\documentclass[10pt]{iopart}

\newcommand{\eff}[0]{{\rm{e}}}

\newcommand{\bra}[1]{\langle #1 |}
\newcommand{\iprod}[2]{\langle #1 |#2 \rangle}
\newcommand{\ket}[1]{| #1 \rangle}

\newcommand{\abs}[1]{\bigl| #1 \bigr|}
\newcommand{\fabs}[1]{\left| #1 \right|}

\newcommand{\ClebschG}[6]{\left[{#1\atop#2}{#3\atop#4}{#5\atop#6}\right]_q}

\newcommand{\F}[6]{F_q\negthinspace\left[#1#3#4#2;#5#6\right]}
\newcommand{\Fun}[6]{F_1\negthinspace\left[#1#3#4#2;#5#6\right]}
\newcommand{\SixJ}[6]{\left\{{#1\atop#2}{#3\atop#4}{#5\atop#6}\right\}_q}
\newcommand{\SixJinline}[6]{\Bigl\{{#1\atop#2}{#3\atop#4}{#5\atop#6}\Bigr\}_q}
\newcommand{\SixJun}[6]{\left\{{#1\atop#2}{#3\atop#4}{#5\atop#6}\right\}_1}

\newcommand{\matr}[1]{\mathbf{#1}}

\usepackage{iopams} 
\usepackage{citesort}
\usepackage{graphicx,color}

\begin{document}

\title[Entanglement spectra of $q$-deformed spin-S VBS states]{Entanglement spectra of $q$-deformed higher spin VBS states}
\author{Raul A~Santos${}^{1}$, Francis N~C~Paraan${}^{1}$, Vladimir E~Korepin${}^{1}$ and Andreas Kl\"umper${}^{2}$}
\address{${}^{1}$ C. N. Yang Institute for Theoretical Physics, State University of New York at Stony Brook, NY 11794-3840, USA}
\address{${}^{2}$ Fachbereich C Physik, Bergische Universit\"at Wuppertal, 42097 Wuppertal, Germany}
\ead{\mailto{santos@insti.physics.sunysb.edu}, \mailto{fparaan@max2.physics.sunysb.edu}, \mailto{korepin@max2.physics.sunysb.edu} and \mailto{kluemper@uni-wuppertal.de}}
\date{\today}
%\preprint{YITP-SB-12-03}

\begin{abstract}
We calculate the reduced density matrix of a block of integer spin-$S$'s in a $q$-deformed valence-bond-solid (VBS) state. This matrix is diagonalized exactly for an infinitely long block in an infinitely long chain. We construct an effective Hamiltonian with the same spectrum as the logarithm of the density matrix. We also derive analytic expressions for the von Neumann and R\'enyi entanglement entropies. For blocks of finite length, we calculate the eigenvalues of the reduced density matrix by perturbation theory and numerical diagonalization. These results enable us to describe the effects of finite-size corrections on the entanglement spectrum and entropy in this generalized VBS model.

\end{abstract}

%03.67.Mn QI Entanglement measures, witnesses, and other characterizations 
%03.65.Ud	QM Entanglement and quantum nonlocality 
%75.10.Jm Quantized spin models, including quantum spin frustration
%75.10.Pq	Spin chain models

\submitto{\JPA}

%\maketitle
\hbadness=10000

\section{Introduction}

We study entanglement in a one-dimensional $q$-deformed valence-bond-solid (VBS) state with a spin-$S$ at each site. This state, which we denote as VBS$_q(S)$, is invariant under the action of the generators of the SU$_q$(2) quantum algebra \cite{jimbo1985,drinfeld1985}. Historically, this algebra featured prominently in the analysis of the quantum sine-Gordon model \cite{kulish1981a} and the anisotropic XXZ Heisenberg chain \cite{reshetikhin1989}. Here we use this formalism to introduce anisotropy into a spin-$S$ VBS state by deforming the usual underlying SU(2) symmetry. In a previous analysis of the $S=1$ VBS$_q(S)$ state we used transfer matrix techniques to calculate the entanglement spectrum and entropy as functions of the parameter $q$ \cite{santos2011}. This investigation and the current one are motivated by two fundamental problems. The first is to quantify the effects of anisotropy on the entanglement in VBS states \cite{santos2011,verstraete2004a,chen2006,li2008b,solanocarrillo2011}. Second, we aim to develop effective thermal models \cite{li2008} that have the same spectrum as the density matrix of a subsystem of a VBS state. This last objective is motivated by our discovery \cite{santos2011} that the deformation parameter $q$ may be interpreted as the temperature of effective models describing the interaction of degrees of freedom on the boundary of the subsystem.

The undeformed VBS state is the ground state of the SU(2) symmetric one-dimensional Affleck-Kennedy-Lieb-Tasaki (AKLT) model \cite{aklt1987,aklt1988}. This model has nearest-neighbour interactions between integer spin-$S$'s. It is described by a Hamiltonian of the form $\mathcal{H} = \sum_i h_{i,i+1}$. The local Hamiltonian $h_{i,i+1}$ is a projector onto the subspace spanned by the $(S+1)$, $(S+2),\dots$, and $2S$-multiplets formed by spins at sites $i$ and $i+1$ \cite{aklt1987,aklt1988,kirillov1989}. Exact results for the block entanglement entropy for the $S=1$ VBS state were obtained in \cite{fan2004,geraedts2010} and for arbitrary integer $S$ in \cite{katsura2007,xu2008}. Generalized VBS states with other symmetries, such as SU($N$) \cite{affleck1991,greiter2007a,greiter2007b}, SO($N$) \cite{tu2008a,tu2008b}, Sp($N$) \cite{schuricht2008}, and supersymmetry (SUSY) \cite{arovas2009,hasebe2011} have also been considered. Entanglement in the SU($N$) \cite{katsura2008,orus2011} and SO($N$) \cite{orus2011b} symmetric VBS state has been studied, but not in the Sp($N$) and SUSY cases. Since these VBS states have matrix product state (MPS) representations \cite{fannes1989,klumper1991,klumper1992,verstraete2006,verstraete2008}, one can obtain the block entanglement in these systems by general transfer matrix techniques \cite{santos2011}.

The anisotropic $q$-deformed generalization of the spin-1 AKLT chain was first considered in \cite{batchelor1990,klumper1991,klumper1992}. The MPS representation of the ground state of the model was constructed in \cite{klumper1991}. This ground state is separated from excited states by a gap \cite{klumper1992}. Hence, the spin-spin correlation functions decay exponentially \cite{klumper1991,klumper1992}. The exact entanglement spectrum of blocks of arbitrary length in the spin-1 VBS$_q(S)$ state was calculated in \cite{santos2011}. The higher integer spin generalization of the $q$-deformed AKLT model was first proposed in \cite{motegi2010,arita2011}, where the spin-spin correlation functions were calculated. To our knowledge, the entanglement spectrum and entropies of VBS$_q(S)$ states for arbitrary integer $S$ have not yet been evaluated.

In this paper, we calculate the entanglement spectrum and entropies of $q$-deformed VBS states with arbitrary integer $S$. Our analytical approach involves transfer matrix methods and the use of $q$-deformed Clebsch-Gordan coefficients and 6$j$ symbols. We begin by constructing the VBS$_q(S)$ state by requiring it to be a ground state of a $q$-deformed spin-$S$ AKLT Hamiltonian. We shall denote this ground state by the state vector $\ket{{\rm{VBS}}_q(S)}$. This state is then partitioned into a block of $\ell$ sequential spins and the environment $E$. The density matrix of the whole ground state is therefore $\rho= \ket{{\rm{VBS}}_q(S)}\bra{{\rm{VBS}}_q(S)}/\iprod{{\rm{VBS}}_q(S)}{{\rm{VBS}}_q(S)}$. We then compute the reduced density matrix $\rho_\ell$ of the block by taking the partial trace of $\rho$ over the environment, $\rho_\ell = \tr_E \rho$. The details of these derivations are provided in \sref{deriv}.

In the double scaling limit of an infinitely long block in an infinitely long chain, we are able to exactly diagonalize the reduced density matrix (\sref{dls1}). We then use the eigenvalues of this matrix to construct the entanglement spectrum of the block (\sref{dls2}). This entanglement spectrum (introduced in \cite{li2008}) enables us to construct an effective Hamiltonian that completely describes the reduced density matrix. The eigenvalues of the reduced density matrix are further used to calculate the R\'enyi and von Neumann entanglement entropies
\begin{eqnarray}
	S_{\rm R}(\alpha) \equiv \frac{\ln\tr\rho_\ell^\alpha}{1-\alpha},\qquad \alpha > 0, \\
	S_{\rm vN} \equiv -\tr(\rho_\ell\ln\rho_\ell) = \lim_{\alpha\to 1} S_{\rm R}(\alpha).
\end{eqnarray}
We thus provide exact measures of entanglement \cite{bennett1996a,bennett1996c,vedral1997,amico2008} in the $\ket{{\rm{VBS}}_q(S)}$ ground state as functions of the deformation parameter $q$ and spin $S$.

We further consider the case of blocks of finite length. We obtain the exact eigenvalues of the reduced density matrix in the isotropic case $q=1$. With this result, we calculate the leading order finite-size corrections to the entanglement spectrum and entropies (\sref{iso}). For a general $q$-deformation, we estimate the eigenvalues of the reduced density matrix in the limit of long but finite blocks by perturbation theory (\sref{arbq}). Furthermore, we numerically investigate the properties of the reduced density matrix of the spin-2 VBS$_q(S)$ state. This result allows us to make general statements about the structure and degeneracy of the entanglement spectrum for blocks of any length.

\section{Model and definitions}\label{deriv}

\subsection{Quantum algebra}\label{quantumalgebra}
Let us denote states of a spin-$S$ at site $i$ by $\ket{S,m}_i$. Here $m \in \{-S,-S+1,\dots,S\}$ is the magnetic quantum number denoting the $z$-component of the spin. The label $S$ of the state $\ket{S,m}_i$ is invariant under the action of $q$-deformed angular momentum operators satisfying the SU$_q$(2) quantum group algebra \cite{drinfeld1985}

\begin{equation}\label{QGalgebra}
[J^{+}_i,J^{-}_i]=[2J_i^z], \qquad [J^{z}_i,J^{\pm}_i]=\pm J^{\pm}_i,
\qquad [n]\equiv\frac{q^{n/2}-q^{-{n/2}}}{q^{1/2}-q^{-{1/2}}}.
\end{equation}
This algebra has two different unitary representations for positive real $q$ and complex $q$ on the unit circle \cite{biedenharn1995}. In this paper we will consider the former case where $q \in \mathbb{R}^+$. The resulting algebra is invariant under the transformation $q\to q^{-1}$ so that we consider further $q\in(0,1]$. The usual SU(2) algebra is recovered at the isotropic point $q=1$, while full deformation occurs in the limit $q\to 0$.  The $q$-number $[n]$ will be used extensively below. 

The analogue of total angular momentum, $\mathbf{J}_{\rm tot}=\mathbf{J}_{1}+\mathbf{J}_{2}$ is realized at the level of operators through the definition
of the coproduct
\begin{eqnarray}
J_{\rm tot}^\pm\equiv q^{-{J_{1}^z}/{2}}\otimes J_{2}^\pm+J_{1}^\pm\otimes q^{{J_{2}^z}/{2}}, \\ 
J_{\rm tot}^z\equiv \mathbb{I}_{1}\otimes J_{2}^z+J_{1}^z\otimes\mathbb{I}_{2}.
\end{eqnarray}
The operators ${J}_{\rm tot}^{z,\pm}$ satisfy the quantum group algebra (\ref{QGalgebra}). A $(2J+1)$-dimensional irreducible representation of $\mathbf{J}_{\rm tot}$ is therefore spanned by the states
\begin{equation}\label{JM}
 \ket{J,m}\equiv\sum_{m_1m_2}\ClebschG{j_1}{m_1}{j_2}{m_2}{J}{m}\ \ket{j_1,m_1}\otimes\ket{j_2,m_2},
\end{equation}
which satisfies
\begin{eqnarray} 
 J_{\rm{tot}}^\pm\ket{J,m} = \sqrt{[J\mp m][J\pm m+1]}\,\ket{J,m\pm1},\\
 J_{\rm{tot}}^z\ket{J,m} = m\,\ket{J,m}.
\end{eqnarray}
These equations define the $q$-deformed Clebsch-Gordan ($q$-CG) coefficients $\ClebschG{J}{m_j}{K}{m_k}{L}{m_l}$ up to a phase \cite{kirillov1989book,biedenharn1995,kassel1995}. 

The $q$-CG coefficients are components of a unitary matrix (a change of basis matrix) and may be chosen to be real. These coefficients vanish if the triangle relation $|j_1-j_2|\le J\le j_1+j_2$ and selection rule $m_1+m_2 = m$ are not satisfied (angular momentum conservation). 
Throughout this paper, the summation indices $m_i$ (lower row of $q$-CG symbols) are understood to run over all values compatible with the corresponding quantum number $j_i$ (upper row of $q$-CG symbols). For example, in \eref{JM} we sum over $m_{i} \in \{-j_i, -j_i+1, \dots ,j_i\}$. Some identities involving the $q$-CG coefficients that we use in the following derivations are collected in \ref{app:qcg}.

\subsection{Matrix product representation}\label{sqvbs}
\begin{figure}[tb]
	\centering
		\includegraphics[width=0.65\linewidth]{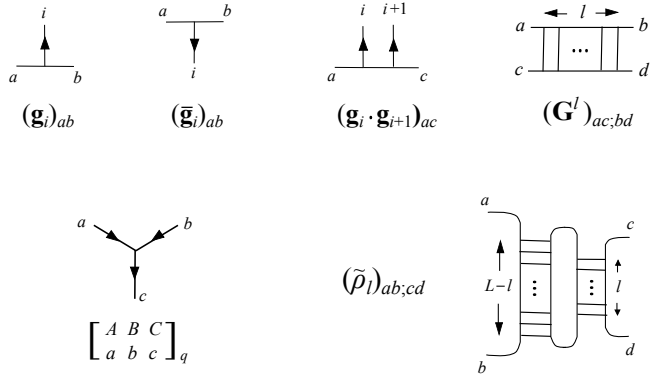}
	\caption{Diagrammatic representations of matrices used in the MPS description of the VBS$_q(S)$ state (upper row). The $q$-deformed Clebsch-Gordan coefficients (lower left) are important in the diagonalization of the transfer matrix. The matrix $\tilde{\rho}_l$ (lower right) is related to the reduced density matrix $\rho_l$ of $l$ sequential spins in a chain of $L$ sites by $\rho_l = \tilde{\rho}_l /\tr \matr{G}^L$.}\label{fig:diag}
\end{figure}

Some of the objects we describe here are conveniently represented as diagrams (\fref{fig:diag}). The manipulation of these diagrams has been useful in the study of entanglement and correlation functions in matrix product states \cite{orus2011,fledderjohann2011}.

\begin{figure}[tb]
	\centering
		\includegraphics[width=0.6\linewidth]{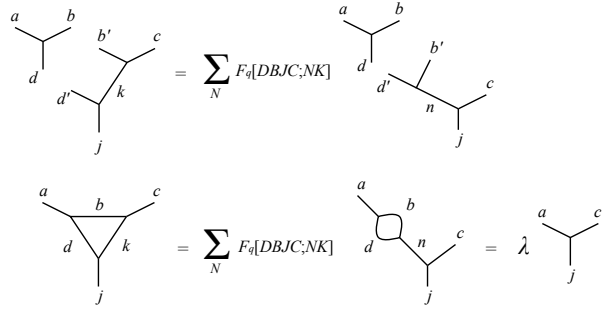}
	\caption{Diagrams representing the contraction of the transfer matrix $\matr{G}$ with a $q$-deformed Clebsch-Gordan coefficient. The $q$-deformed $F$-matrix is defined according to the upper diagram in which the leg labeled by $b'$ is shifted. The lower diagram represents the eigenvalue equation \eref{eigenfinal}. In these diagrams internal lowercase indices are summed over.}\label{fig:identity}
\end{figure}

Let us now write down the MPS representation of the VBS$_q(S)$ state. For a periodic chain of $L$ spins we have 
\begin{equation}\label{mps}
	\ket{{\rm{VBS}}_q(S)} = \tr\bigl(\matr{g}_1\bdot\matr{g}_2\bdot\dots\bdot\matr{g}_L\bigr),
\end{equation}
where $\matr{g}_i$ are $(S+1)\times(S+1)$ matrices. The trace here is done over the auxiliary matrix space. The elements of $\matr{g}_i$ and its dual $\bar{\matr{g}}_i$ are state vectors:

\begin{eqnarray}\label{def_g}
 (\matr{g}_i)_{ab}=\sum_{m}{\ClebschG{\frac{S}{2}}{a}{\frac{S}{2}}{-b}{S}{m}}(-1)^bq^{-b/2}\ \ket{S,m}_i, \\
	(\bar{\matr{g}}_i)_{ab}=\sum_{m}\ClebschG{\frac{S}{2}}{a}{\frac{S}{2}}{-b}{S}{m}(-1)^bq^{-b/2}\ \bra{S,m}_i.
\end{eqnarray}
The $\ket{{\rm{VBS}}_q(S)}$ state \eref{mps} is annihilated by the $q$-deformed AKLT Hamiltonian (periodic boundary conditions)
\begin{equation}\label{hamiltonian}
	\mathcal{H} = \sum_{i=1}^L h_{i,i+1} = \sum_{i=1}^L \sum_{s=S+1}^{2S} \Pi_s(i,i+1),
	%\sum_i h_{i,i+1}\ket{{\rm{VBS}}_q(S)} = \sum_i \sum_{s=S+1}^{2S} \Pi_s(i,i+1)\,\ket{\rm{VBS}_q(S)} =0,
\end{equation}
where $\Pi_s(i,i+1)$ is a projector onto the subspace spanned by the $q$-deformed $s$-multiplet formed by spins at $i$ and $i+1$. To prove this, we look at the overlap between the two states
\begin{eqnarray}
	 \fl (\matr{g}_i\bdot\matr{g}_{i+1})_{ac}= \sum_{bm'm}(-1)^{c-b}\mbox{$q^{\frac{1}{2}(b-c)}$} \ClebschG{\frac{S}{2}}{a}{\frac{S}{2}}{b}{S}{m'}\ClebschG{\frac{S}{2}}{-b}{\frac{S}{2}}{-c}{S}{m} \,\ket{S,m'}_i\otimes\ket{S,m}_{i+1}, \\
	 \fl \ket{J,m}=\sum_{m_1m_2}\ClebschG{j_1}{m_1}{j_2}{m_2}{J}{m}\ \ket{j_1,m_1}_i\otimes\ket{j_2,m_2}_{i+1}.
\end{eqnarray}
Since the states $\{\ket{j,m}_i\}$ are orthonormal to each other, we obtain
\begin{equation}
	\fl \bra{J,m}(\matr{g}_i\bdot\matr{g}_{i+1})_{ac} =\negthinspace \negthinspace \sum_{bm_1m_2}(-1)^{c-b}\mbox{$q^{\frac{1}{2}(b-c)}$}\ClebschG{\frac{S}{2}}{a}{\frac{S}{2}}{b}{S}{m_1}\ClebschG{\frac{S}{2}}{-b}{\frac{S}{2}}{-c}{S}{m_2}\ClebschG{S}{m_1}{S}{m_2}{J}{m}\negthinspace.
\end{equation}
Using an identity \eref{idfinal} derived in \ref{app:fmatrix} yields
\begin{equation}
	\fl\bra{J,m}(\matr{g}_i\bdot\matr{g}_{i+1})_{ac} = (-1)^{c-S/2}\mbox{$q^{-c/2}$}\sqrt{\frac{[2S+1]}{[S+1]}}\,\F{S}{\mbox{$\frac{S}{2}$}}{\mbox{$\frac{S}{2}$}}{J}{\mbox{$\frac{S}{2}$}}{S}\ClebschG{{\mbox{$\frac{S}{2}$}}}{a}{{\mbox{$\frac{S}{2}$}}}{-c}{J}{m}.\label{overlap}
\end{equation}
The elements $F_q[DBJC;NK]$ of the $q$-deformed $F$-matrix are defined diagrammatically in \fref{fig:identity}. The $q$-CG coefficient in the overlap \eref{overlap} vanishes when $J>\frac{S}{2}+\frac{S}{2}$ proving that $h_{i,i+1}\ket{{\rm{VBS}}_q(S)} = 0$. Furthermore, $h_{i,i+1}$ has nonnegative eigenvalues because it is a sum of projectors. Thus, $\ket{{\rm{VBS}}_q(S)}$ is a ground state of the Hamiltonian \eref{hamiltonian}. 

\subsection{Transfer matrix}\label{transfer}
We now construct a transfer matrix ${\matr{G}}$ that is defined in terms of $\matr{g}$ and $\bar{\matr{g}}$ by $(\matr{G})_{aa';bb'}=(\bar{\matr{g}})_{ab}(\matr{g})_{a'b'}$. Explicitly, its elements are
\begin{equation}
 (\matr{G})_{aa';bb'}=\sum_{m'}\ClebschG{\frac{S}{2}}{a}{\frac{S}{2}}{-b}{S}{m'}\ClebschG{\frac{S}{2}}{a'}{\frac{S}{2}}{-b'}{S}{m'}(-1)^{b+b'}q^{-({b+b'})/{2}}.
\end{equation}
This transfer matrix is an important object that appears in the calculation of correlation functions and the reduced density matrix of MPS. Let us diagonalize this matrix through an approach based on the $q$-deformed $F$-matrix (\ref{app:fmatrix}). As depicted in \fref{fig:identity}, we construct the eigenvalue equation $(\matr{G})_{aa';bb'}e_{bb'}=\lambda e_{aa'}$ using the $q$-CG coefficients as an ansatz for the elements of the eigenvector $e_{aa'}$. The resulting equation is
\begin{equation}
\fl (\matr{G})_{aa';bb'}e_{bb'}= \sum_{bb'm'}\ClebschG{\frac{S}{2}}{a}{\frac{S}{2}}{-b}{S}{m'}\ClebschG{\frac{S}{2}}{b}{j}{m}{\frac{S}{2}}{b'}\ClebschG{\frac{S}{2}}{a'}{\frac{S}{2}}{-b'}{S}{m'}(-1)^{b+b'}q^{-({b+b'})/{2}}.
\end{equation}

\noindent Transposing columns in the third $q$-CG coefficient in order to match the identity (\ref{idfinal}) leads to

\begin{eqnarray}\nonumber
\fl (\matr{G})_{aa';bb'}e_{bb'} &= \sum_{bb'm'}\ClebschG{\frac{S}{2}}{a}{\frac{S}{2}}{b}{S}{m'}\ClebschG{\frac{S}{2}}{-b}{j}{m}{\frac{S}{2}}{b'}\ClebschG{S}{m'}{\frac{S}{2}}{b'}{\frac{S}{2}}{a'}(-1)^{-\frac{S}{2}-b}q^{{b}/{2}}\sqrt{\frac{[2S+1]}{[S+1]}}\\
\fl &=(-1)^{-S}\frac{[2S+1]}{[S+1]}\F{S}{j}{\mbox{$\frac{S}{2}$}}{\mbox{$\frac{S}{2}$}}{\mbox{$\frac{S}{2}$}}{\mbox{$\frac{S}{2}$}}\ClebschG{\frac{S}{2}}{a}{j}{m}{\frac{S}{2}}{a'}=\lambda e_{aa'}.\label{eigenfinal}
\end{eqnarray}

We see that the elements of the eigenvectors of $\matr{G}$ are $e_{aa'}=(e_{jm})_{aa'}=\ClebschG{{S}/{2}}{a}{j}{m}{{S}/{2}}{a'}$. A suitable similarity transformation on $\matr{G}$ gives the orthonormal set of eigenvectors 
\begin{equation}
	(\hat{e}_{jm})_{aa'}=\ClebschG{\frac{S}{2}}{-a}{\frac{S}{2}}{a'}{j}{m}.\label{evecG}
\end{equation}
To obtain this set we transposed the middle and last rows of $(e_{jm})_{aa'}$ and considered the orthogonality relation \eref{rows}. The eigenvalues associated with these eigenvectors are
\begin{eqnarray}
	\lambda_{j} &=(-1)^{-S}\frac{[2S+1]}{[S+1]}\F{S}{j}{\mbox{$\frac{S}{2}$}}{\mbox{$\frac{S}{2}$}}{\mbox{$\frac{S}{2}$}}{\mbox{$\frac{S}{2}$}}, \\
	&= (-1)^{j+S}[2S+1]\SixJ{S}{j}{\mbox{$\frac{S}{2}$}}{\mbox{$\frac{S}{2}$}}{\mbox{$\frac{S}{2}$}}{\mbox{$\frac{S}{2}$}}, \label{lambda2}
	%\frac{(-1)^{S+J}\,[2S+1]\,[S]!^2}{[S-J]!\,[S+J+1]!},\qquad \mbox{with }[x]! \equiv [x][x-1]!\ \mbox{\ and \,} [0]!\equiv 1. \label{eigenG}
	%\fl&=\frac{(-1)^S}{[2S]!}\frac{(-1)^J[S]!^2\,[2S+1]!}{[S-J]!\,[S+J+1]!},\qquad \mbox{with }[x]! \equiv [x][x-1]!\ \mbox{\ and \,} [0]!\equiv 1. \label{eigenG}
	%\lambda_J&=(-1)^{-S}\frac{[2S+1]}{[S+1]}\F{S}{J}{\mbox{$\frac{S}{2}$}}{\mbox{$\frac{S}{2}$}}{\mbox{$\frac{S}{2}$}}{\mbox{$\frac{S}{2}$}} \nonumber \\
\end{eqnarray}
where the $q$-deformed 6$j$ symbol in the second line is defined in \eref{Fqto6j}. The eigenvalue $\lambda_{j}$ is $(2j+1)$-fold degenerate and $0\le j \le S$. The absolute value of $\lambda_j$ decreases with increasing $j$. These expressions match the results of \cite{motegi2010,arita2011} (except for a multiplicative constant).

\subsection{Reduced Density Matrix}\label{rdm}
In this subsection we calculate the reduced density matrix $\rho_\ell$ of $\ell$ sequential spins in a chain of infinite length $L\to\infty$. Using the formalism developed in \cite{santos2011} for matrix product states, we obtain
\begin{equation}\label{rdmell1}
	(\rho_\ell)_{ab;cd} = \frac{1}{\tr\matr{G}^L}\sum_{a'b'} \bigl(\matr{G}^{L-\ell}\bigr)_{aa';bb'}\bigl(\matr{G}^{\ell}\bigr)_{a'c;b'd}.
\end{equation}
Integer powers of the transfer matrix $\matr{G}$ may be written as
\begin{equation}
	\matr{G}^n = \sum_j \lambda_j^n \matr{P\negthinspace}_{j},
\end{equation}
where $\matr{P\negthinspace}_j$ is a projection matrix onto the subspace spanned by the eigenvectors $\hat{e}_{jm}$ of $\matr{G}$. Since the dominant (largest magnitude) eigenvalue of $\matr{G}$ is $\lambda_{j=0}$, large integer powers of $\matr{G}$ simplify to $\matr{G}^n \to \lambda_0^n \matr{P\negthinspace}_0$ as $n\to\infty$. Thus, in the limit of infinite chains $L\to \infty$ the reduced density matrix \eref{rdmell1} simplifies to
\begin{equation}
	(\rho_\ell)_{ab;cd} = \sum_{a'b'} (\matr{P\negthinspace}_0)_{aa';bb'}\sum_{j=0}^S \frac{\lambda_j^\ell}{\lambda_0^\ell} (\matr{P\negthinspace}_j)_{a'c;b'd}.
\end{equation}
Constructing the projectors $\matr{P\negthinspace}_j$ from the eigenvectors of $\matr{G}$ yields
\begin{equation}
\fl	 (\rho_\ell)_{ab;cd} = \frac{q^{-(a+b)/2}(-1)^{a+b+S}}{[S+1]}\sum_{j=0}^S\frac{\lambda_j^\ell}{\lambda_0^\ell} \sum_{m=-j}^j \ClebschG{\frac{S}{2}}{-a}{\frac{S}{2}}{c}{j}{m}\ClebschG{\frac{S}{2}}{-b}{\frac{S}{2}}{d}{j}{m}, \\
\end{equation}
where the projector $\matr{P\negthinspace}_0$ is simplified by the explicit formula \cite{biedenharn1995}
\begin{equation}
\ClebschG{\frac{S}{2}}{-a}{\frac{S}{2}}{a}{0}{0}=\frac{(-1)^{a+S/2}q^{-a/2}}{\sqrt{[S+1]}}.
\end{equation}
We can further express the reduced density matrix as a sum of tensor products by defining the $(S+1)\times(S+1)$ matrix
\begin{equation}\label{qmatrix}
	\bigl(\matr{Q}_{jm}\bigr)_{ac} \equiv \frac{(-1)^{a+S/2}q^{-a/2}}{\sqrt{[S+1]}}\ClebschG{\frac{S}{2}}{-a}{\frac{S}{2}}{c}{j}{m} \delta_{m,c-a}.
\end{equation}
The Kronecker delta here enforces the triangle relation. Finally, making the necessary substitutions gives 
\begin{equation}\label{rdmcompact}
\rho_\ell = \sum_{j=0}^S\frac{\lambda_j^\ell}{\lambda_0^\ell} \sum_{m=-j}^{j} \matr{Q}_{jm}\otimes\matr{Q}_{jm}.
\end{equation}

\section{Double scaling limit}\label{dls}
\subsection{Eigenvalues of reduced density matrix}\label{dls1}
In the double scaling limit, we consider infinitely long blocks and take $\ell\to\infty$. The reduced density matrix $\rho_\ell$ \eref{rdmcompact} simplifies to a tensor product of diagonal matrices $\rho_\infty = \matr{Q}_{00}\otimes\matr{Q}_{00}$. Explicitly, we have:
\begin{equation}
(\rho_\infty)_{ab;cd}= \frac{(-1)^{a+b+S}q^{-(a+b)/2}}{[S+1]} \ClebschG{\frac{S}{2}}{-a}{\frac{S}{2}}{a}{0}{0}\ClebschG{\frac{S}{2}}{-b}{\frac{S}{2}}{b}{0}{0} \delta_{ac}\delta_{bd}.
\end{equation}
Remembering that $-{S}/{2}\leq a,b,c,d\leq {S}/{2}$, with integer steps, we arrive at the final expression for the reduced density 
matrix, 
\begin{equation}\label{rdmdls}
 (\rho_\infty)_{ab;cd}=\frac{q^{-(a+b)}}{[S+1]^2}\delta_{ac}\delta_{bd}.
\end{equation}
From this expression we can compute all eigenvalues of $\rho_\infty$. For example, in the case of a $q$-deformed spin-$2$ VBS state we have
\begin{equation}
	\rho_\infty = \frac{1}{(q+1+q^{-1})^2}\ \left(\begin{array}{ccc} q^{-1}&0&0\\0&1&0 \\ 0&0&q \end{array}\right)\otimes\left(\begin{array}{ccc} q^{-1}&0&0\\0&1&0 \\ 0&0&q \end{array}\right).
\end{equation} 
The nine eigenvalues of this matrix are proportional to $\{q^{2},q,q,1,1,1,q^{-1},q^{-1},q^{-2}\}$.

\subsection{Entanglement spectrum and entropy}\label{dls2}
We now write $\rho_\infty = \rme^{-\beta H_\eff}/\tr \rme^{-\beta H_\eff}$, where $H_\eff$ is an effective Hamiltonian and $1/\beta$ an effective temperature. The eigenvalues of the Hamiltonian $H_\eff$ constitute the entanglement spectrum of the block \cite{li2008}. The tensor product form of $\rho_\infty$ \eref{rdmdls} yields the simple paramagnetic model
\begin{eqnarray}
	-\beta H_\eff = 
	-\beta \bigl( H_\eff^{(1)} + H_\eff^{(2)}\bigr) \equiv
	 \beta h ( {S}_1^z + {S}_2^z\bigr), \qquad \mathbf{S}_{i}^2 = \mbox{$\frac{S}{2}(\frac{S}{2}+1)$}, \\
	 \beta h = \left| \ln q\right|.
\end{eqnarray}
Here $h$ is the magnitude of an effective magnetic field along the $z$-axis,  while $\mathbf{S}_{i}$ are spin-$S/2$ operators of the undeformed SU(2) algebra. We can thus identify $\left| \ln q\right|$ as the ratio $h/T_\eff$ between the magnitude of the magnetic field and effective temperature $T_\eff$. We observe that the spectrum of $H_\eff^{(i)}$ consists of $S+1$ equidistant energy levels. Thus, in the limit $S\to\infty$ the entanglement spectrum of the block is equal to the energy spectrum of two harmonic oscillators with frequency $\omega$ (with an $S$-dependent energy shift). This frequency is related to the deformation parameter through $\left| \ln q\right| = \omega/T_\eff$.

In this effective picture, the isotropic case $q=1$ corresponds to infinite temperature or zero field strength. The block is therefore maximally mixed. The reduced density matrix $\rho_\infty$ has $(S+1)^2$ nonzero identical eigenvalues. In the opposite limit $q\to 0^+$ the effective model corresponds to zero temperature or infinite field magnitude. Hence, the block is in a pure state with zero entanglement.

\begin{figure}[tb]
	\centering
		\includegraphics[width=0.45\linewidth]{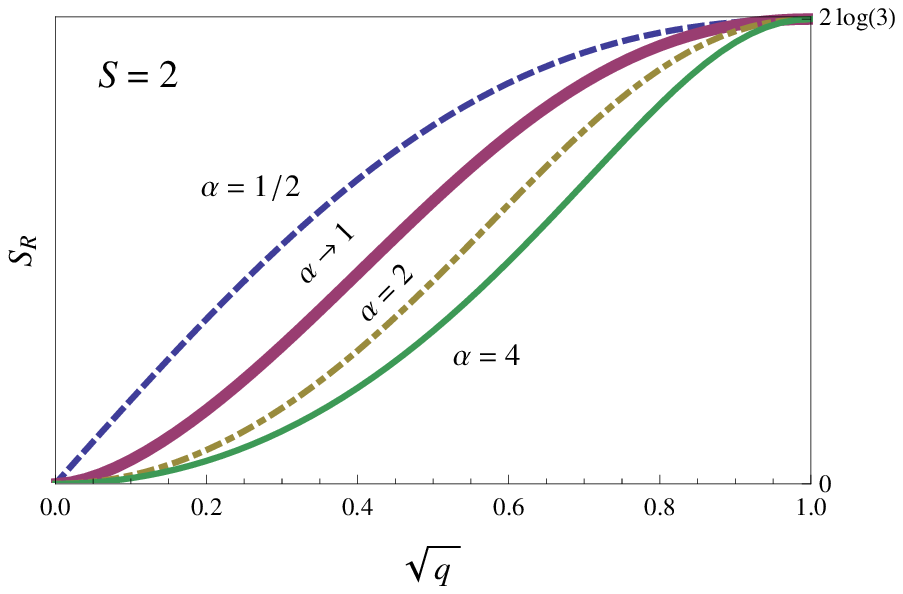}
		\includegraphics[width=0.45\linewidth]{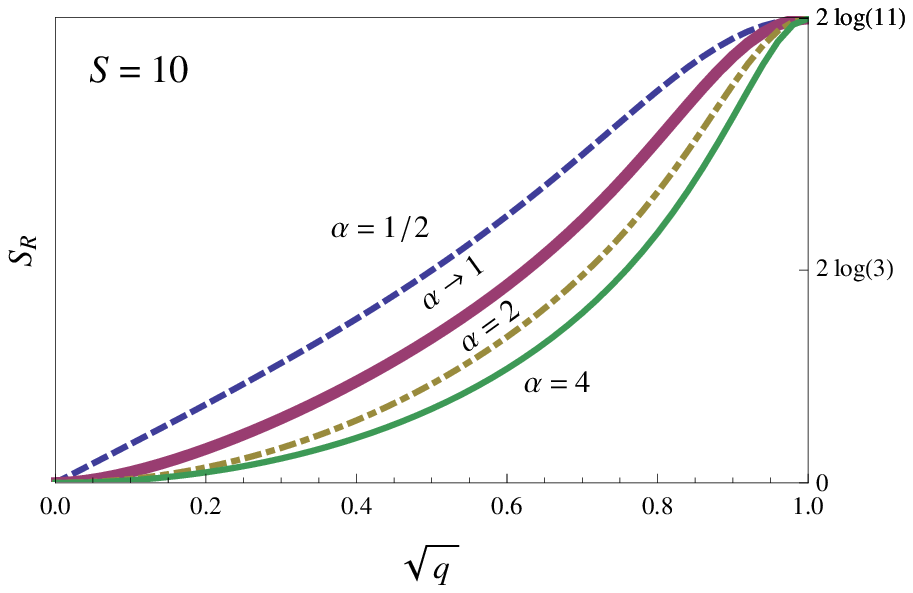}
	\caption{The R\'enyi entropy $S_{\rm{R}}(\alpha)$ of a $q$-deformed spin-$S$ VBS state vanishes in the limit $q\to0^+$. At the isotropic point $q=1$, large blocks are maximally entangled $S_{\rm R}(\alpha) = 2\ln(S+1)$. The von Neumann entropy is obtained in the limit $\alpha\to 1$ (bold line).}\label{fig:s10}
\end{figure}

We use the eigenvalues of the reduced density matrix \eref{rdmdls} to compute the R\'enyi entropy 
\begin{eqnarray}
S_{\rm R}(\alpha)&=\frac{\ln\tr\rho^\alpha}{1-\alpha}%=2\frac{\ln\tr\rho_0^\alpha}{1-\alpha}
 =\frac{2}{1-\alpha}\ln\left\{\frac{q^\frac{\alpha(S+1)}{2}-q^{-\frac{\alpha(S+1)}{2}}}{q^\frac{\alpha}{2}-q^{-\frac{\alpha}{2}}}\ \frac{1}{[S+1]^\alpha}\right\}, \nonumber \\ &=\frac{2}{1-\alpha}\ln\left\{\frac{q^\frac{\alpha(S+1)}{2}-q^{-\frac{\alpha(S+1)}{2}}}{q^\frac{\alpha}{2}-q^{-\frac{\alpha}{2}}}\left(\frac{q^\frac{1}{2}-q^{-\frac{1}{2}}}{q^\frac{S+1}{2}-q^{-\frac{S+1}{2}}}\right)^\alpha\right\}.\label{srenyi}
\end{eqnarray}
This is an exact expression in the double scaling limit (infinite block). Taking the limit $\alpha\rightarrow 1$ gives the von Neumann entropy

\begin{eqnarray}
 \fl S_{\rm vN} &= 2\ln\bigl([S+1]\bigr) +\left\{\frac{q^\frac{1}{2}+q^{-\frac{1}{2}}}{q^\frac{1}{2}-q^{-\frac{1}{2}}}
-(S+1)\frac{q^\frac{S+1}{2}+q^{-\frac{S+1}{2}}}{q^\frac{S+1}{2}-q^{-\frac{S+1}{2}}}\right\}\ln q, \nonumber \\
\fl &= 2 \ln\left\{\frac{q^\frac{S+1}{2}-q^{-\frac{S+1}{2}}}{q^\frac{1}{2}-q^{-\frac{1}{2}}}\right\} +\left\{\frac{q^\frac{1}{2}+q^{-\frac{1}{2}}}{q^\frac{1}{2}-q^{-\frac{1}{2}}}
-(S+1)\frac{q^\frac{S+1}{2}+q^{-\frac{S+1}{2}}}{q^\frac{S+1}{2}-q^{-\frac{S+1}{2}}}\right\}\ln q.\label{svn}
\end{eqnarray}
In order to recover previous results for the spin-1 case \cite{santos2011}, we have to rescale $q\rightarrow q^2$ in \eref{srenyi} and \eref{svn}. This transformation is necessary because of the different definition \eref{QGalgebra} used here for the deformation parameter $q$.

These entanglement entropies are graphed in \fref{fig:s10} as functions of the parameter $q$ for the cases $S=2,10$. At the isotropic point $q = 1$, the entanglement entropies simplify to
\begin{equation}
	S_{\rm{R}}(\alpha) = S_{\rm{vN}} = 2\ln (S+1),\qquad q=1.
\end{equation}
We thus recover previous results \cite{katsura2007,xu2008} for isotropic spin-$S$ VBS states. In the limit of full deformation $q\to0^+$ the entanglement entropy for any spin $S$ vanishes.

Finally, let us consider the case of very high spin at fixed $0<q<1$. Taking the limit $S\to \infty$ in \eref{srenyi} and \eref{svn} gives
\begin{eqnarray}
	S_{\rm R}(\alpha) = \frac{2}{1-\alpha}\ln\biggl\{\frac{\bigl(q^{-1/2}-q^{1/2}\bigr)^\alpha}{q^{-\alpha/2}-q^{\alpha/2}}\biggr\},\\
	S_{\rm vN} = 2\ln\biggl(\frac{1}{q^{-1/2}-q^{1/2}}\biggr) + \biggl(\frac{q^{1/2}+q^{-1/2}}{q^{1/2}-q^{-1/2}}\biggr)\ln q,\qquad S\to\infty.
\end{eqnarray}
We find that the entanglement entropy is bounded for any $q$-deformed AKLT chain of arbitrary spin $S$. It diverges only at the isotropic point $q=1$. 

\section{Finite-size corrections}\label{finite}
We now look at the case of finite blocks in an infinite chain. The reduced density matrix \eref{rdmcompact} of a block of $\ell$ spins may be written as
\begin{equation}
\rho_\ell = \rho_\infty + \sum_{j=1}^S\frac{\lambda_j^\ell}{\lambda_0^\ell}\sum_{m=-j}^{j} \matr{Q}_{jm}\otimes\matr{Q}_{jm}.
\end{equation}
Let us express this matrix in the basis of the following vectors:
\begin{equation}
	(\matr{v}_{\negthinspace JM})_{ab} = \frac{(-1)^{-(J+b)}q^{b/2}}{\sqrt{[2J+1]}}\,\ClebschG{S/2}{a}{S/2}{-b}{J}{M},
\end{equation}
with corresponding dual
\begin{equation}
	(\bar{\matr{v}}_{\negthinspace JM})_{ab} = {(-1)^{J+b}}q^{-b/2}\sqrt{{[2J+1]}}\,\ClebschG{S/2}{a}{S/2}{-b}{J}{M}.
\end{equation}
This choice of basis allows us to use identity \eref{idfinal} and get
\begin{eqnarray}
\fl (\rho_\ell)_{J'M',JM} = \frac{(-1)^{J'-J}}{[S+1]^2}\sqrt{\frac{[2J'+1]}{[2J+1]}}\Biggl(1 + \sum_{j=1}^S [2j+1]\frac{\lambda_j^\ell}{\lambda_0^\ell}\,\F{j}{J}{\mbox{$\frac{S}{2}$}}{\mbox{$\frac{S}{2}$}}{\mbox{$\frac{S}{2}$}}{\mbox{$\frac{S}{2}$}}\Biggr)\nonumber \\ \times  \delta_{M'M}\sum_{ab}q^{-(a+b)}\ClebschG{S/2}{a}{S/2}{-b}{J'}{M}\ClebschG{S/2}{a}{S/2}{-b}{J}{M}.\label{matelemrho}
\end{eqnarray}
We find that the reduced density matrix can be decomposed into sectors labelled by the quantum number $M\in[-S,S]$. Each sector is represented by an $(S+1-|M|)\times(S+1-|M|)$ matrix. These matrices are diagonal in the $\{\matr{v}_{\negthinspace JM}\}$ basis only when $q=1$. We treat this isotropic case analytically in \sref{iso}. For the general case $0<q<1$, we calculate the eigenvalues of $\rho_\ell$ using first-order perturbation theory and compare this approximation with numerical results in \sref{arbq}.

\subsection{Isotropic case}\label{iso}
When $q=1$, using the orthogonality condition \eref{rows} in \eref{matelemrho} yields the exact $(2J+1)$-fold degenerate eigenvalues
\begin{eqnarray}
	p_{JM} &= \frac{1}{(S+1)^2}\biggl(1 + \sum_{j=1}^S (2j+1)\frac{\lambda_j^\ell}{\lambda_0^\ell}\,\Fun{j}{J}{\mbox{$\frac{S}{2}$}}{\mbox{$\frac{S}{2}$}}{\mbox{$\frac{S}{2}$}}{\mbox{$\frac{S}{2}$}}\biggr), \\
	&=\frac{1}{(S+1)^2} + \frac{(-1)^{J+S}}{S+1}\sum_{j=1}^S (-1)^j(2j+1)\,\frac{\lambda_j^\ell}{\lambda_0^\ell}\,\SixJun{j}{J}{\mbox{$\frac{S}{2}$}}{\mbox{$\frac{S}{2}$}}{\mbox{$\frac{S}{2}$}}{\mbox{$\frac{S}{2}$}}.\label{pJ}
\end{eqnarray}
with $0\le J \le S$ and $M\in \{-J, -J+1,\dots,J\}$. For instance, taking $S=2$ gives the exact eigenvalues
\begin{eqnarray}
	p_{00} = \mbox{$\frac{1}{9}$}\left(1 + 3(-2)^{-\ell}+5(10)^{-\ell} \right),\qquad&\mbox{(degeneracy 1)}, \\
	p_{1M} = \mbox{$\frac{1}{9}$}\left(1 + \mbox{$\frac{3}{2}$}(-2)^{-\ell} -\mbox{$\frac{5}{2}$}(10)^{-\ell} \right),\qquad&\mbox{(degeneracy 3)}, \\
	p_{2M} = \mbox{$\frac{1}{9}$}\left(1 - \mbox{$\frac{3}{2}$}(-2)^{-\ell} +\mbox{$\frac{1}{2}$}(10)^{-\ell} \right),\qquad&\mbox{(degeneracy 5)}.
\end{eqnarray}
The formula \eref{pJ} reproduces the results of \cite{katsura2007,xu2008} for undeformed spin-$S$ AKLT chains obtained from the Schwinger boson representation of the VBS state. Our approach, however, emphasizes the role of $6j$ symbols in determining finite-size effects on entanglement in these states. 
Additionally, this result solves a recursive formula in \cite{katsura2007,xu2008} for the coefficients in the sums for the eigenvalues $p_{JM}$. 

The leading finite-size correction to the eigenvalue $p_{JM}$ is proportional to the exponential factor $(\lambda_1/\lambda_0)^\ell \equiv (-1)^\ell\rme^{-\ell/\xi}$. Using the formula
\begin{equation}\label{ident4}
	\SixJun{S}{j}{\mbox{$\frac{S}{2}$}}{\mbox{$\frac{S}{2}$}}{\mbox{$\frac{S}{2}$}}{\mbox{$\frac{S}{2}$}} = \frac{(S!)^2}{(S-j)!(S+j+1)!},
\end{equation}
gives the characteristic length of decay $\xi = 1/\ln\bigl((S+2)/S\bigr)$. This length is equal to the correlation length of the spin-spin correlation functions in the spin-$S$ VBS state \cite{arita2011}.

Let us construct an effective Hamiltonian for long blocks $1\ll \ell < \infty$ in the isotropic case. Considering only the leading-order correction to the eigenvalues \eref{pJ} gives
\begin{equation}
	\fl p_{JM} \approx \frac{1}{(S+1)^2}\left\{1 - \frac{3}{S(S+2)}\biggl(\frac{-S}{S+2}\biggr)^\ell\bigl(2J(J+1)-S(S+2)\bigr)\right\}.
\end{equation}
Since the reduced density matrix is diagonal in the $\{\matr{v}_{\negthinspace JM}\}$ basis, we can write the effective Hamiltonian $H_\eff$ as
\begin{eqnarray}
	 -\beta H_\eff &\approx \ln\left\{1 - \frac{3}{S(S+2)}\biggl(\frac{-S}{S+2}\biggr)^\ell\bigl(2J(J+1)-S(S+2)\bigr)\right\}, \nonumber \\
	&\approx - \frac{12}{S(S+2)}\biggl(\frac{-S}{S+2}\biggr)^\ell\times \mbox{$\frac{1}{2}$}\bigl\{J(J+1)-S(\mbox{$\frac{S}{2}$}+1)\bigr\}.
\end{eqnarray}
This expression is valid for $3S^\ell(S+2)^{-\ell} \ll 1$. If we define an undeformed spin-$S$ operator $\mathbf{J}\equiv\mathbf{S}_1+\mathbf{S}_\ell$ as the sum of two spin-$\frac{S}{2}$ operators $\mathbf{S}_{1}$ and $\mathbf{S}_{\ell}$ on the block boundaries, we obtain the Heisenberg model
\begin{equation}
\beta H_\eff = \gamma\negthinspace\left(S,\ell\right)(-1)^{\ell}\,\mathbf{S}_1\bdot\mathbf{S}_\ell, \qquad 
	\gamma\negthinspace\left(S,\ell\right) = \frac{12}{S(S+2)}\biggl(\frac{S}{S+2}\biggr)^\ell. \label{isotropicH}
\end{equation}

We can identify $T_\eff = 1/\gamma(S,\ell)$ as an effective temperature that depends on the length of the block. The double scaling limit $\ell \to\infty$ therefore corresponds to a maximally mixed state (infinite temperature). In this interpretation, we observe that the sign of the coupling strength changes with block length (alternation between ferromagnetic and antiferromagnetic interactions). This implies that the dominant eigenvalue of the reduced density matrix alternates between the $p_{00}$ singlet (even $\ell$) and $p_{SM}$ multiplet (odd $\ell$). 

Let us now consider the entanglement entropy of a block consisting of a single spin ($\ell =1$) for the case $q=1$. The eigenvalues of the reduced density matrix may be written as
\begin{equation}
\fl p_{JM}(\ell=1) = \frac{1}{(S+1)^2} + \frac{(-1)^{J+S}}{\mbox{$\SixJun{S}{0}{S/2}{S/2}{S/2}{S/2}$}}  \sum_{j=1}^S\frac{2j+1}{S+1}\SixJun{S}{j}{\mbox{$\frac{S}{2}$}}{\mbox{$\frac{S}{2}$}}{\mbox{$\frac{S}{2}$}}{\mbox{$\frac{S}{2}$}}\SixJun{j}{J}{\mbox{$\frac{S}{2}$}}{\mbox{$\frac{S}{2}$}}{\mbox{$\frac{S}{2}$}}{\mbox{$\frac{S}{2}$}}.
\end{equation}
Making use of the identity \eref{ident4} and
\begin{eqnarray}
	\SixJun{S}{0}{\mbox{$\frac{S}{2}$}}{\mbox{$\frac{S}{2}$}}{\mbox{$\frac{S}{2}$}}{\mbox{$\frac{S}{2}$}} \SixJun{J}{0}{\mbox{$\frac{S}{2}$}}{\mbox{$\frac{S}{2}$}}{\mbox{$\frac{S}{2}$}}{\mbox{$\frac{S}{2}$}}= \frac{(-1)^{J+S}}{(S+1)^2}, \\
	\sum_{j=0}^S (2j+1)\SixJun{S}{j}{\mbox{$\frac{S}{2}$}}{\mbox{$\frac{S}{2}$}}{\mbox{$\frac{S}{2}$}}{\mbox{$\frac{S}{2}$}}\SixJun{j}{J}{\mbox{$\frac{S}{2}$}}{\mbox{$\frac{S}{2}$}}{\mbox{$\frac{S}{2}$}}{\mbox{$\frac{S}{2}$}} =\frac{\delta_{SJ}}{2S+1},
\end{eqnarray}
gives the desired result
\begin{equation}
	p_{JM}(\ell=1) = \frac{\delta_{SJ}}{2S+1},\qquad q=1.
\end{equation}
Thus, the single-site reduced density matrix has $(2S+1)$ nonzero identical eigenvalues. This result proves that the block is a uniform mixture of the $(2S+1)$ states of a single spin-$S$ as expected. The entanglement entropy in this case is $S_{\rm{R}}(\alpha)= S_{\rm{vN}} = \ln (2S+1)$. 

For long blocks satisfying $\ell \gg \xi$, the leading nonvanishing correction to the entanglement entropy is proportional to $(\lambda_1/\lambda_0)^{2\ell}$. The approximate R\'enyi entropy in this case is 
\begin{equation}
	 S_{\rm R}(\alpha)\approx 2\ln(S+1) -\frac{3\alpha}{2}\left(\frac{S}{S+2}\right)^{2\ell}S(S+1)(S+2).
\end{equation} 
Finite-size corrections to the von Neumann entropy can be obtained from this result by taking the limit $\alpha\to1$.

\begin{figure}[tb]
	\centering
		\includegraphics[width=0.4\linewidth]{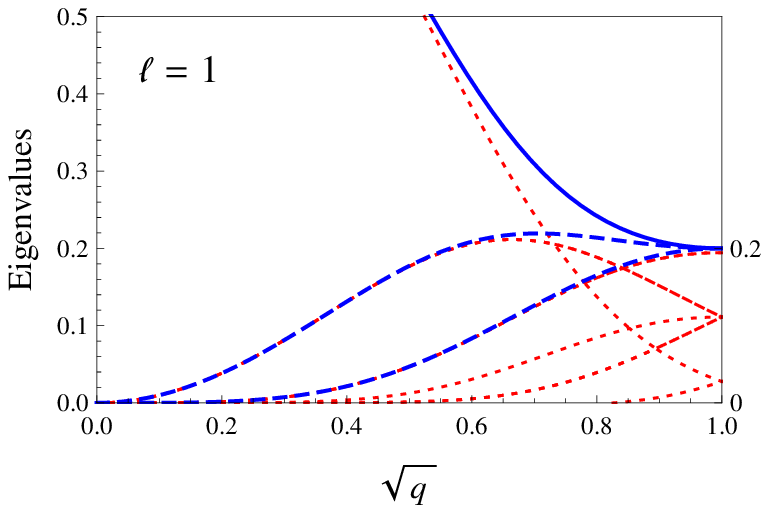}
		\includegraphics[width=0.4\linewidth]{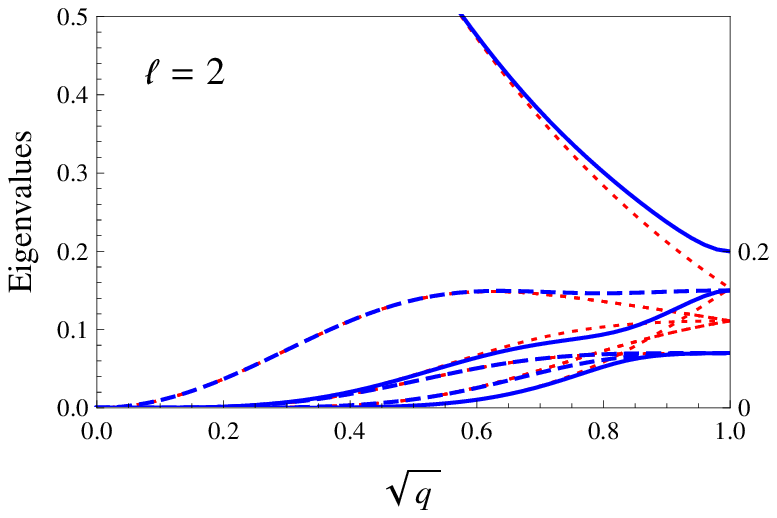}
		\includegraphics[width=0.4\linewidth]{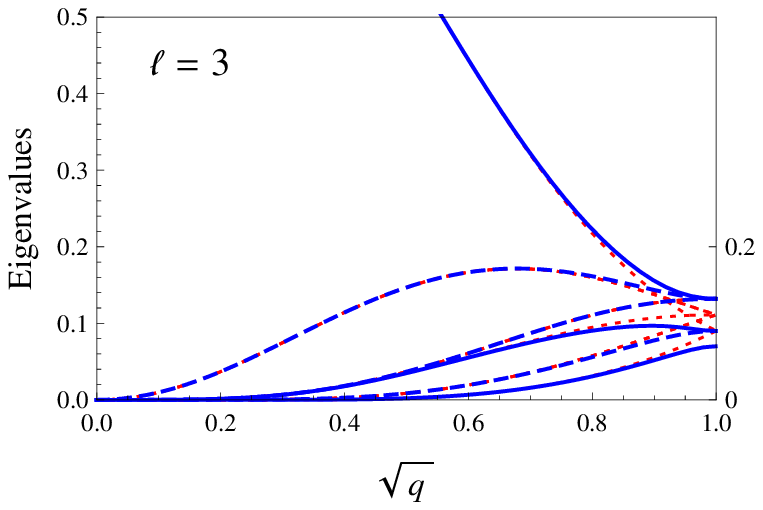}
		\includegraphics[width=0.4\linewidth]{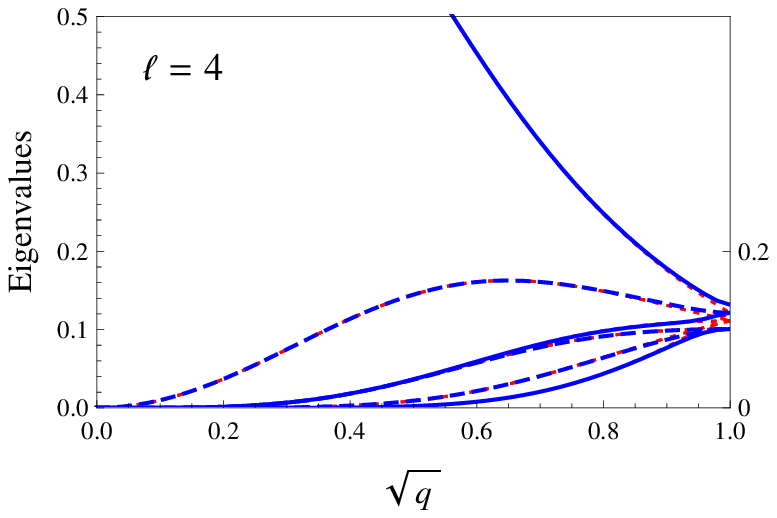}
	\caption{The eigenvalues of the reduced density matrix of a block of $\ell$ spins in a spin-2 VBS$_q$ state (solid and dashed blue lines) are compared to the perturbation result (red dotted lines). Solid blue lines denote nondegenerate eigenvalues while dashed blue lines denote doubly degenerate ones. The dominant eigenvalue approaches unity as $q\to 0$. For $\ell =1$ four eigenvalues are zero for all $q$.}\label{fig:spin2}
\end{figure}

\subsection{Anisotropic case}\label{arbq}
For arbitrary values of $q$, the dominant characteristic length of finite-size corrections generalizes to $\xi =  1/\ln\bigl([S+2]/[S]\bigr)$. That is, we have $\abs{\lambda_1/\lambda_0} = [S]/[S+2] <1$. For blocks of length $\ell \gg \xi$, we may therefore approximate the reduced density matrix as
\begin{equation}
	\rho_\ell \approx \matr{Q}_{00}\otimes\matr{Q}_{00} + \frac{\lambda_1^\ell}{\lambda_0^\ell}\sum_{m=-1}^{1} \matr{Q}_{1m}\otimes\matr{Q}_{1m}.
\end{equation}
We have already determined that $\matr{Q}_{00}$ is diagonal with nondegenerate eigenvalues \eref{qmatrix}. This means that first-order perturbation theory within each sector of the preceding equation involves only the diagonal elements of $\matr{Q}_{1m}$. From \eref{qmatrix} we know that only $\matr{Q}_{10}$ has nonzero diagonal elements and hence we obtain the approximate eigenvalues 
\begin{equation}
	p_{\mu\nu} =p_{\nu\mu} \approx \frac{q^{-(\mu+\nu)}}{[S+1]^2}\left(1 + [3]\,\frac{\lambda_1^\ell}{\lambda_0^\ell}\ClebschG{\mbox{$\frac{S}{2}$}}{\mu}{1}{0}{\mbox{$\frac{S}{2}$}}{\mu}\ClebschG{\mbox{$\frac{S}{2}$}}{\nu}{1}{0}{\mbox{$\frac{S}{2}$}}{\nu} \right).\label{perturb}
\end{equation} 
The labels $\mu$ and $\nu$ are quantum numbers that run from $-\frac{S}{2}$ to $\frac{S}{2}$ with integer steps. The second term in \eref{perturb} involving the $q$-CG coefficients may be evaluated explicitly with the identity \cite{biedenharn1995}
\begin{equation}
\fl	\ClebschG{\mbox{$\frac{S}{2}$}}{\mu}{1}{0}{\mbox{$\frac{S}{2}$}}{\mu} = \frac{q^{-\mu/2}}{\sqrt{[S][S+2]}}\,\Bigl\{q^{\mbox{\scriptsize $\frac{1}{2}$}(1+S/2)}\left[\mbox{$\frac{S}{2}$}+\mu\right] - q^{-\mbox{\scriptsize$\frac{1}{2}$}(1+S/2)}\left[\mbox{$\frac{S}{2}$}-\mu\right]\Bigr\}.
\end{equation}
These approximate eigenvalues are compared to exact numerical results for the spin-2 case in \fref{fig:spin2}. We observe a rapid improvement in the accuracy of the perturbation result with increasing block length $\ell$. Furthermore, these numerical results reveal how $q$-deformation modifies the degeneracy of the entanglement spectrum by breaking the multiplet structure present in the isotropic case.

\section{Conclusions}
We exactly calculated the reduced density matrix of $q$-deformed VBS states with arbitrary integer spin-$S$ in the double scaling limit. We discovered that the entanglement spectrum corresponds to a thermal ensemble of two spin-$S$/2's in a uniform magnetic field $h$. The deformation parameter $q$ enters this picture as the effective field divided by the temperature $\fabs{\ln q} = h/T_\eff$. In this infinite block limit, we also derived exact expressions for the R\'enyi and von Neumann entropies as functions of the deformation parameter $q$ and spin $S$. We used this result to demonstrate that entanglement in a VBS$_q(S)$ state is monotonically reduced by increasing anisotropy. 

Furthermore, we obtained the exact reduced density matrix of finite blocks in an infinite chain. We diagonalized this matrix for the isotropic case $q=1$ and obtained its exact spectrum in terms of $6j$ symbols. We found that degenerate eigenvalues of the reduced density matrix are grouped into multiplets. For long but finite blocks, we constructed an approximate effective Hamiltonian in the undeformed case. This effective model consists of two spin-$S$/2's with a Heisenberg interaction. The coupling parameter of this Heisenberg Hamiltonian alternates between ferromagnetic and antiferromagnetic depending on the parity of $\ell$.

We also studied the general case of arbitrary deformation $q$ and block length $\ell$. We made approximations for the eigenvalues of the reduced density matrix of long blocks using first-order perturbation theory. Finally, we numerically investigated the spectrum of the reduced density matrix of finite blocks in a deformed spin-2 VBS$_q(S)$ state. In this case we discovered that $q$-deformation partially breaks the degeneracy of eigenvalues within each multiplet. 

We have mentioned in the introduction that generalized VBS states have been constructed with various symmetries. The dimension of the MPS representation of these states depends on the particular representation chosen for the given symmetry group. Let us consider a $D\times D$ MPS. On general grounds one can 
state that the rank of the reduced density matrix is less than or equal to $D^2$ regardless of the number of spins in the block \cite{santos2011}. Thus, the entanglement entropy is always bounded according to $S \le 2\ln D$. For example, the isotropic SU($N$) VBS state is described by an $N \times N$ $\matr{g}$ matrix. This entropy bound is saturated in the double scaling limit where it is known that the entanglement entropy is $S=2 \ln N$ \cite{katsura2007,xu2008}. In this case, we expect anisotropy by $q$-deformation to reduce entanglement entropy in the SU($N$) VBS state as we proved here (\fref{fig:s10}).  

\ack
The authors are grateful for the hospitality of the Simons Center for Geometry and Physics (Stony Brook, New York). R.S., F.P., and V.K. acknowledge support by the National Science Foundation through Grant No. DMS-0905744. R.S. is supported by a Fulbright-CONICYT grant. A.K. thanks the C.~N.~Yang Institute for Theoretical Physics for its hospitality.

\appendix
\section{Identities for $q$-CG coefficients}\label{app:qcg}

Among the key properties of the $q$-CG coefficients that we use above are the orthogonality relations

\begin{eqnarray}
 %\sum_{J,m}\ClebschG{j_1}{m_1}{j_2}{m_2}{J}{m}\ClebschG{j_1'}{m_1'}{j_2}{m_2'}{J}{m}=\delta_{m_1,m_1'}\delta_{m_2,m_2'}, \quad \mbox{or rows,}\\
 \label{column}\sum_{Jm}\ClebschG{j_1}{m_1}{j_2}{m_2}{J}{m}\ClebschG{j_1}{m_1'}{j_2}{m_2'}{J}{m}=\delta_{m_1m_1'}\delta_{m_2m_2'}, \quad \mbox{(columns),}\\
 \label{row}\sum_{m_1m_2}\ClebschG{j_1}{m_1}{j_2}{m_2}{J}{m}\ClebschG{j_1}{m_1}{j_2}{m_2}{J'}{m'}=\delta_{JJ'}\delta_{mm'}, \quad \mbox{(rows).} \label{rows}
\end{eqnarray}
We also make much use of the following identities involving column transpositions: 
\small
\begin{eqnarray}\label{sym1}
 \ClebschG{j_1}{m_1}{j_2}{m_2}{J}{m}&=&(-1)^{j_1-J+m_2}q^{{-m_2}/{2}}\sqrt{\frac{[2J+1]}{[2j_1+1]}}\ClebschG{J}{m}{j_2}{-m_2}{j_1}{m_1},\\
 \ClebschG{j_1}{m_1}{j_2}{m_2}{J}{m}&=&\ClebschG{j_2}{-m_2}{j_1}{-m_1}{J}{-m},\\
 \ClebschG{j_1}{m_1}{j_2}{m_2}{J}{m}&=&(-1)^{J-j_2-m_1}q^{{m_1}/{2}}\sqrt{\frac{[2J+1]}{[2j_2+1]}}\ClebschG{j_1}{-m_1}{J}{m}{j_2}{m_2}.
\end{eqnarray}
\normalsize

\section{$q$-deformed $F$-matrix and 6$j$ symbols}\label{app:fmatrix}
The equation for the lower diagram given in \fref{fig:identity} reads

\begin{eqnarray}\label{id1}
\fl \sum_{abcdk}\ClebschG{A}{a}{B}{b}{D}{d}\ClebschG{D}{d}{K}{k}{J}{j}\ClebschG{B}{b}{C}{c}{K}{k}\ket{A,a}\otimes\ket{C,c} =\nonumber \\ \fl \qquad
\sum_{N}\sum_{abcdn}\F{D}{C}{B}{J}{N}{K}\ClebschG{A}{a}{B}{b}{D}{d}\ClebschG{D}{d}{B}{b}{N}{n}\ClebschG{N}{n}{C}{c}{J}{j}\ket{A,a}\otimes\ket{C,c}.
%\fl \sum_{abckd}\ClebschG{A}{a}{B}{b}{D}{d}\ClebschG{D}{d}{K}{k}{J}{j}\ClebschG{B}{b}{C}{c}{K}{k}\ket{A,a}\ket{C,c}= \nonumber \\ \fl \qquad
%\sum_n\SixJ{D}{C}{B}{J}{N}{K}\Biggl(\sum_{abndc}\ClebschG{A}{a}{B}{b}{D}{d}\ClebschG{D}{d}{B}{b}{N}{n}\ClebschG{N}{n}{C}{c}{J}{j}\ket{A,a}\ket{C,c}\Biggr).
\end{eqnarray}
Using the identity \eref{sym1} in the righthand side of \eref{id1} and applying the orthogonality condition \eref{rows} to evaluate the sum gives

\begin{eqnarray}\label{idfinal}
\fl\sum_{bdk}\ClebschG{A}{a}{B}{b}{D}{d}\ClebschG{B}{-b}{C}{c}{K}{k}\ClebschG{D}{d}{K}{k}{J}{j}\!\!\!\!(-1)^{-b}q^{{b}/{2}}  \nonumber \\
%=(-1)^{A-D}\sqrt{\frac{[2D+1]}{[2A+1]}}\SixJ{D}{C}{B}{J}{A}{K}\ClebschG{A}{a}{C}{c}{J}{j}, \\
= (-1)^{A-D}\sqrt{\frac{[2D+1]}{[2A+1]}}\,\F{D}{C}{B}{J}{A}{K}\ClebschG{A}{a}{C}{c}{J}{j}, \\
= (-1)^{A+B+C+J}\sqrt{[2D+1][2K+1]}\SixJ{D}{C}{B}{J}{A}{K}\ClebschG{A}{a}{C}{c}{J}{j}. \label{idfinal6j}
%\fl\sum_{a,b,c,k,d}\ClebschG{A}{a}{B}{b}{D}{d}\ClebschG{B}{b}{C}{c}{K}{k}\ClebschG{D}{d}{K}{k}{J}{j}\!\!\!\!(-1)^{-b}q^{{b}/{2}} = \nonumber \\
%(-1)^{A-D}\sqrt{\frac{[2A+1]}{[2D+1]}}\SixJ{D}{C}{B}{J}{A}{K}\ClebschG{A}{a}{C}{c}{J}{j}.
\end{eqnarray}
Here $\SixJinline{D}{C}{B}{J}{A}{K}$ is the $q$-deformed 6$j$ symbol. It is related to the elements of the $q$-deformed $F$-matrix by \cite{biedenharn1995}
\begin{equation}\label{Fqto6j}
\fl \F{D}{C}{B}{J}{A}{K} = (-1)^{D+B+J+C}\sqrt{[2K+1][2A+1]}\SixJ{D}{C}{B}{J}{A}{K}.
\end{equation}

\section*{References}
%\bibliographystyle{iopart-num}
%\bibliography{sqvbsbib}

\providecommand{\bibNYu}{N Yu} \providecommand{\bibYuA}{Yu A}
\providecommand{\newblock}{}

\end{document}